# Some Astrophysical Implications of Compact Extra Dimensions

Michael J. Longo

*University of Michigan, Department of Physics, Ann Arbor MI 48109-1120*

There have been many suggestions that there are extra spatial dimensions "outside" of our normal (3+1)-dimensional space. Generally it is assumed that electromagnetic and hadronic fields are restricted to the normal dimensions, while gravity can extend into the extra dimensions. Thus a hadron or lepton is confined in a potential well, and excited states should exist. I explore the possibility that ordinary hadrons and leptons have excited states ("compactons"), some of which may have electromagnetism and possibly gravity confined to the <u>extra</u> spatial dimensions. Some of these may be long-lived. If sufficiently long-lived, relic compactons may exist as dark matter. Black holes may be the source of new particles with bizarre properties that appear as UHE cosmic rays.

## 1. Introduction

There has been considerable discussion on the possibility of extra spatial dimensions, which are compactified on some distance scale $R$ [1]. In gravity-mediated theories the associated energy scale is $M \sim R^{-1} \sim 10^{13}$ GeV. In weak-scale compactification theories [2] the energy scale is ~1 TeV. It has been suggested that the compactification of the extra dimensions may occur on scales ≲ 1 mm with energy scales also in the TeV range [3]. In all theories, because of the success of the Standard Model, electroweak and hadronic fields must be confined to the 3 ordinary spatial dimensions, at least down to scales approaching the Planck scale. Gravity, however, can penetrate into the extra dimensions, and the strength of the gravitational interaction increases dramatically on scales ≲ $R$.

In this paper I assume there are extra compactified dimensions with $R$ somewhere between the Planck scale and 1 mm. I speculate on some of the possible ramifications, particularly in astrophysics.

## 2. Some astrophysical implications of large-scale compactification

Given a theory with a built-in length scale $R$ and the greatly increased strength of the gravitational force for distances $\lesssim R$, it is plausible (though not necessarily compelling) that black holes with dimensions $\sim R$ and masses $\sim R^{-1}$ would be formed in the early universe. If $R \sim 0.1$ mm, this would correspond to black holes with masses comparable to that of the moon. Thus much of the dark matter in the universe might now consist of black holes with mass smaller than that of the Moon. These would be very difficult to observe. If their average mass density near the solar system were comparable to that for ordinary matter, the chance of a close encounter with Earth would be comparable to that for the Earth to encounter a Moon-sized normal object. If $R \sim 1$ fm, the corresponding black hole mass is $\sim 10^{12}$ kg. Such less massive versions would be more plentiful, but correspondingly more difficult to observe. They might, for example, pass through the Earth with very little effect. In (3+1)-dimensional general relativity, black holes with mass $\lesssim 10^{12}$ kg would evaporate due to Hawking radiation with lifetimes much less than the age of the universe. However, the much greater strength of gravity in (4+n) dimensions for distances $\lesssim R$ would slow their evaporation, so primordial black holes with masses $\sim 10^4$ kg might still exist.[4]

If gravity were significantly stronger at short distance scales, the evolution of the universe would be profoundly affected while its size was $\lesssim R$. Depending on $R$, this could cause difficulties with inflation. In effect we are translating our previous ignorance of what happens below the Planck scale to the compactification scale.

## 3. Do ordinary particles have counterparts with different compactified dimensions?

Given the existence of compactified extra dimensions, we can picture an ordinary hadron or lepton as having the electroweak and hadronic parts of its wave function confined almost completely to ordinary 3-dimensional space. In other words there is very little penetration into the compactified dimensions. (The success of the standard model up to energy scales $\sim 1$ TeV proves that it is valid to distance scales $\sim 10^{-19}$ m.) The particle is effectively confined to a potential well $\leq 10^{-19}$ m wide in the compactified dimensions.

If ordinary quantum mechanics is applicable in the compactified dimensions, there must be excited states of ordinary hadrons and leptons that correspond to the excited states of the classic problem of a particle in a potential well. (We assume, of course, that the "wall" does not have zero thickness [5].) Some of these might have quite different symmetries than the ground states. As discussed below, they are likely to have bizarre properties compared to ordinary hadrons. They might, for example, be the supersymmetric partners of the ordinary fermions. For definiteness, I shall refer to these states as *compactons*. Some compacton states should be at least quasi-stable, similar to the situation in alpha decay where there is only a large potential well separating the states. This could give a huge range of lifetimes for different compacton states, but the lowest energy states are likely to be quite long-lived due to the large potential barrier separating the normal state and compacton counterpart. Compacton masses could be modest (and possibly negative!), but they would be very difficult or impossible to produce at accelerators. If stable or very long-lived, they might exist as relics of the big bang, or they may be produced in the present universe near or even inside the Schwarzchild radius of black holes. If they can occasionally escape, they might be the source of UHE cosmic rays or other cosmic ray phenomena, which are otherwise very difficult to explain.

It is difficult to even speculate on compacton properties. An assumption of theories with large extra compactified dimensions is that the standard-model fields are confined to the 3 ordinary spatial dimensions for distances $>>R$, while gravity can penetrate into the extra dimensions. Correspondingly, there could be compacton states with strong and electroweak fields not confined to ordinary dimensions as depicted in Figure 1. Fig. 1(a) shows the electric field around a normal particle with the *x*-axis a normal dimension and the *y*-axis a compactified dimension. Fig. 1(b) illustrates a possible compacton state with the electric field extending into the "extra" dimensions and restricted from normal dimensions. Such a particle, though charged, would have almost no electromagnetic interactions with normal particles. Formally, a normal state can be transformed into a new compacton state by rotating the gauge fields in (3+*n*) spatial dimensions from an ordinary spatial dimensions into (normally) compactified ones.

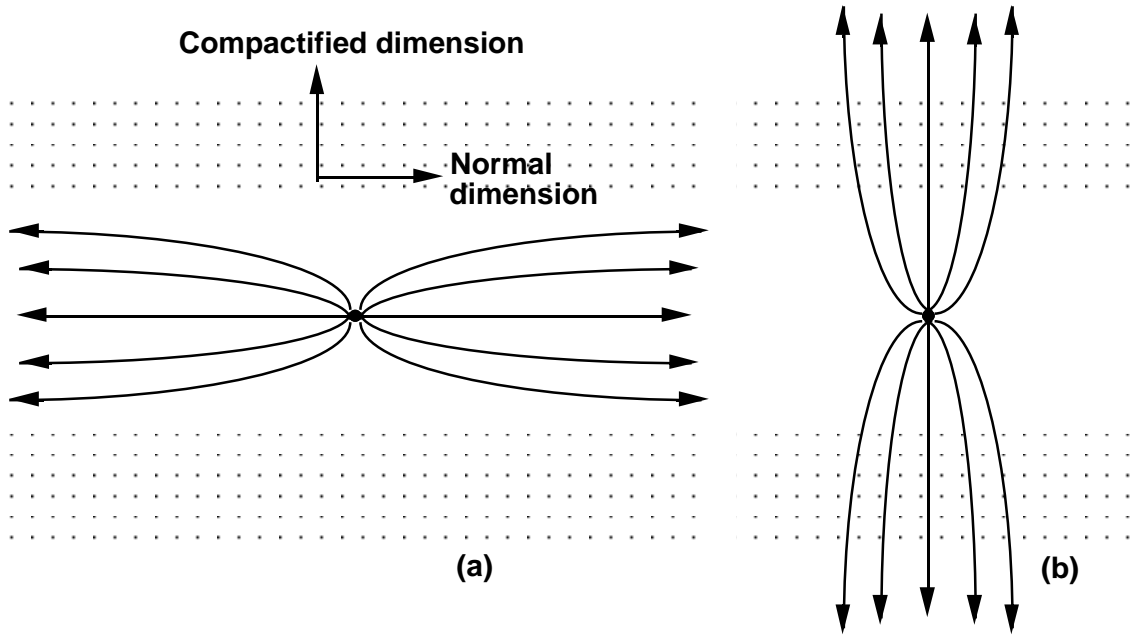

Figure 1   (a) A normal charged particle with its electric field confined to ordinary spatial dimensions;  (b) A compacton state with the electric field confined to extra dimensions.   Such a state would have little electromagnetic interaction with normal particles, but might have normal gravitational interactions.

States might also exist with gravity confined to the extra dimensions [though these could not be formed by a simple rotation in $(3+n)$ space].  It is thus plausible that the gravitational interactions of some compacton states are quite different than for ordinary particles, while their other interactions may be similar.  They could, for example, have no or very weak gravitational interaction with ordinary matter or possibly even repulsive gravity.  Thus once they are formed in a black hole they could easily escape.  Anti-gravity would help to explain the acceleration mechanism for UHE cosmic rays.

The possibility of particles with intrinsic angular momentum components in the extra dimensions gives even more degrees of freedom.

## 4.   The ultimate symmetry

In a sense this is the ultimate symmetry – symmetry with respect to compactification. – Some compacton states might have electromagnetic and/or hadronic fields that are compactified in "ordinary" dimensions and long-range in what we take to be compactified dimensions.  I propose in particular that compacton states exist for which gravity is confined to the <u>extra</u> dimensions for distances $\gg R$, just as the standard model fields are confined to normal dimensions for ordinary particles. This automatically gives these compactons no or very weak large-scale gravitational interactions with ordinary matter, so that they would be able to escape a black hole's gravity. Given this scenario, production of compactons in a black hole is possible. For example, protons falling into a rotating black hole along the axis of rotation will gain large amounts of kinetic energy and occasionally collide with matter falling through the disc from the other side. If the black hole is rotating, the singularity is thought to be disc-shaped, so that the protons can approach arbitrarily close without actually entering the singularity [6], and thus they could collide with almost arbitrarily large energies.

As described above, rotating the electromagnetic field (in $3+n$ dimensions) out of the normal dimensions into the extra ones could produce a compacton with no long-range electromagnetic field in normal dimensions [7]. Relic long-lived compacton states with no electromagnetic or hadronic fields in normal dimensions could be a major component of dark matter. Another possibility is that compacton pairs could be produced from black holes by a mechanism similar to Hawking radiation.

## 5. Compactons as UHE cosmic rays

If symmetry with respect to compactification is a (badly broken) symmetry of nature, it is plausible that compactons can have a "gravitational charge", just as ordinary particles have electromagnetic or color charge. A neutral or negative gravitational charge (mass) would facilitate their escape from black holes.[‡1] The gravitational field of the black from distances arbitrarily close to the singularity they can occasionally be ejected with extremely high energies. If their electromagnetic fields are compactified in ordinary dimensions, their interactions with the microwave background radiation would be quite weak, while their hadronic interactions could be similar to ordinary hadrons. Thus they

---

[‡1] The repulsive gravity of long-lived negative mass compactons surviving from the big bang or produced later in black holes thus could conceivably account for the cosmological constant.

hole would accelerate a negative mass compacton. Because the compactons could come would be able to traverse intergalactic distances without significant energy loss and interact in our atmosphere much like ordinary hadrons. They might account for cosmic rays observed with energies >$10^{20}$ eV [8] that appear to interact in the atmosphere like ordinary hadrons, but apparently do not interact as expected with the intervening cosmic microwave background radiation. Compacton states with different hadronic interactions might also account for other exotic phenomena in cosmic rays such as Centauro events [9].

## 6. Conclusions

I have explored the provocative possibility that ordinary hadrons and leptons have counterparts with gravity and/or electromagnetic fields confined to the extra dimensions rather than the ordinary 3 spatial dimensions. These should exist as excited states of normal particles whose standard model fields are confined in (3+$n$) dimensions. Some of these states may be long-lived.

Compacton states with standard model fields confined to the extra dimensions would have little interaction with ordinary matter. If their long-range gravitational fields are confined to extra dimensions, their gravitational interaction with ordinary matter could be very weak or repulsive. These states, which I call compactons, may be produced in black holes. They may have quite bizarre properties such as negative or zero gravitational mass that would allow them to escape from a black hole. Black holes might then be the source of UHE cosmic rays.

In the early universe the increased strength of the gravitational force for distances below $R$ would naturally lead to the formation of black holes with dimensions ~$R$ and masses ~$R^{-1}$. Due to the increased strength of the gravitational interaction on scales <<$R$, these black holes are unlikely to have evaporated.